\documentclass[preprint,aps,superscriptaddress]{revtex4}%
\usepackage{amsfonts}
\usepackage{amsmath}
\usepackage{amssymb}
\usepackage[dvips]{graphicx}%
\begin{document}
\preprint{HEP/123-qed}
\title{A Tonks Giradeau Gas in the Presence of a Local Potential}
\author{Hao Fu}
\affiliation{Michigan Center for Theoretical Physics, FOCUS Center, and Physics Department,
University of Michigan, Ann Arbor, Michigan 48109-1040}
\author{A. G. Rojo}
\affiliation{Department of Physics, Oakland University, Rochester, MI 48309}


\begin{abstract}
The physics of a Tonks-Giradeau Gas in the presence of a local potential is
studied. In order to evaluate the single particle density matrix (SPDM) of the
many-body ground state, the Wiger-Jordan transformation is used. The
eigenvector with the largest eigenvalue of the SPDM corresponds to the
"Bose-Einstein Condensate"(BEC) State. We find that the "BEC" state density at
the positon of the local potential decreases, as expected, in the case of a
repulsive potential. For an attractive potential, it decreases or increases
depending on the strength of the potential. The superfluidity of this system
is investigated both numerically and perturbatively. An experimental method
for detecting the effect of an impurity in a Tonks-Giradueau gas is discussed.

\end{abstract}
\date{Jan 08 2006}
\maketitle

\section{\bigskip Introduction}

The effect of impurity and disorder on many body system has been one of the
main themes of condensed matter physics\cite{Anderson}. It plays an important
role in our understanding of phenomena such as superconductivity,
superfluidity, and Kondo physics. The rapid progress in ultra-cold atom
experiments provides a unique platform for the study of such many-body
systems. In these experiments, the external potential, the dimensionality, and
the interaction can be tuned using external fields. Fascinating many body
phenomena have already been realized in experiments \cite{becexp}, including
recent realization of a Tonks-Girardeau Gas \cite{Tonksexp}.

A natural extension of this line of research is the study of the effect of
impurities in these many body systems, both theoretically \cite{IT}, and
experimentally \cite{IE}. Here we study the Tonks-Girardeau gas in the
presence of a local potential. A Tonks-Giradeau gas is a quantum gas
consisting of hard core bosons. This system can be mapped to a free fermion
system, allowing for an exact solution. The study of the effect of an impurity
on such an exactly solvable system serves two purposes: first, it shows, in
its simplest form, the interplay between interaction, superfluidity, and
impurity. Second, the theoretical predictions can be compared readily with
experimental observables.

\section{Model of the Tonks-Girardeau Gas with an impurity}

The Hamiltonian we use to describe such a system is of the tight binding type
\begin{equation}
H=t\sum_{i}\left(  a_{i}^{\dag}a_{i+1}+a_{i+1}^{\dag}a_{i}\right)
+ua_{0}^{\dag}a_{0} \label{Boson}%
\end{equation}
where $a_{i}^{\dag}$ and $a_{i}$ are the bosonic operators that create and
annihilate a particle at site $i$, respectively. They obey the commutation
relation $\left[  a_{i},a_{j}^{\dag}\right]  =\delta_{ij}$. The particle
interaction is included by imposing $a_{i}^{2}=\left(  a_{i}^{\dag}\right)
^{2}=0$. The quantity $t,$ which is generally negative, is the hopping
amplitude between nearest neighbors and $u$ characterizes the local potential
strength at site zero. The behavior of the many body system is determined by
the ratio of $u/t$. \ When $\left\vert u/t\right\vert <1$, the kinetic parts
dominate and the particles distribute uniformly to lower the kinetic energy.
On the other hand, when $\left\vert u/t\right\vert >1,$ the local potential
becomes important. In this regime, local charge fluctuations are suppressed
and \ the coherence among particles is degraded. This degradation is a main
interest of this paper. We numerically evaluate the SPDM of the many-body
ground state, and compute the effect of a local potential (impurity) on the
spectrum of the sSPDM, on the "BEC" wavefunction, and on the superfluidity of
the many-body ground state.

In order to solve for the ground state of the system, the
Hamiltonian(\ref{Boson}) can be mapped to an non-interacting fermion
Hamiltonian via the Wigner-Jordan Transformation%
\begin{equation}
a_{i}=%
{\textstyle\prod\limits_{\alpha=1}^{i-1}}
e^{i\pi c_{\alpha}^{\dag}c_{\alpha}}c_{i},a_{i}^{\dag}=c_{i}^{\dag}%
{\textstyle\prod\limits_{\alpha=1}^{i-1}}
e^{-i\pi c_{\alpha}^{\dag}c_{\alpha}} \label{WJ}%
\end{equation}
Here the $c_{i}^{\dag}$ and $c_{i}$ are fermionic operators, which satisfy the
anticommunication relation $\left\{  c_{i},c_{j}^{\dag}\right\}  =\delta_{ij}%
$. Under this transformation, it is straightforward to show that the
Hamiltonian[\ref{Boson}] can be written in the following form
\begin{equation}
H=t\sum_{i}\left(  c_{i}^{\dag}c_{i+1}+c_{i+1}^{\dag}c_{i}\right)
+uc_{0}^{\dag}c_{0} \label{Fermion}%
\end{equation}
The ground state $\left\vert G_{F}\right\rangle $ of this Hamiltonian
corresponds to the well-known Fermi sea, in which fermions fill the single
particle levels up to the Fermi surface. The corresponding ground state
$\left\vert G_{B}\right\rangle $ of the boson Hamiltonian(\ref{Boson}) can be
obtained from $\left\vert G_{F}\right\rangle $ by symmetrizing the
corresponding many body fermionic wave function.

\section{\bigskip Condensate Fraction and Condensate Wavefunction}

We seek the one particle density matrix \cite{Penrose}, which is the
expectation value of $a_{i}^{\dag}a_{j}$ in the ground state.
\begin{align}
\rho_{ij}  &  ={}\left\langle G_{B}\right\vert a_{i}^{\dag}a_{j}\left\vert
G_{B}\right\rangle \nonumber\\
&  =\left\langle {}G_{F}\right\vert c_{i}^{\dag}%
{\textstyle\prod\limits_{\alpha=1}^{i-1}}
e^{-i\pi c_{\alpha}^{\dag}c_{\alpha}}%
{\textstyle\prod\limits_{\alpha=1}^{j-1}}
e^{i\pi c_{\alpha}^{\dag}c_{\alpha}}c_{j}\left\vert G_{F}\right\rangle
\label{DM}%
\end{align}

The diagonal part of the density matrix, $\rho_{ii}$, gives the density
$\left\langle n_{i}\right\rangle ={}\left\langle G_{B}\right\vert a_{i}^{\dag
}a_{i}\left\vert G_{B}\right\rangle $ at site $i$, while the off diagonal part
gives the coherence in the many body ground state for different sites. In the
uniform case, the coherence $\rho_{ij}$ depends only on the difference,
$\left\vert i-j\right\vert $. In the presence of a local potential, however,
the coherence depends on both $i$ and $j$ separately. \ In addition to that,
$\rho_{ij}$ is much smaller than its uniform counterpart when $i$ and $j$ are
on different sides of the local potential. Diagonalizing the density matrix
gives a set of eigenvectors corresponding to the reduced one-particle states.
The corresponding eigenvalues represent the occupation probability for the
corresponding reduced one particle states. The one with a significant large
eigenvalue is the "BEC" State \cite{Penrose} (Note that in 1D, there should be
no Bose Einstein Condensation in the thermodynamical limit. In the particular
case of the Tonks-Girardeau case, the particles that condense into a single
state are calculated to be order of $\sqrt{N}$ \cite{Lenard}. \ However, this
state, compared with other states, is the only state that is significantly
occupied. With this clarification, we will call that state the BEC state and
the corresponding occupation the condensate fraction).

The evaluation of this density matrix is not simple even when the ground state
is known. Here we adopt the technique developed by M. Rigol and A. Muramatsu
\cite{Rigol}. This method allows us to calculate a system with up to a
thousand lattice sites. We focus on the low occupation limit only, which
corresponds to a continuum case. We have verified that 9 particles in 100
lattice sites already converges very well to the continuum case. In the
following calculation, unless otherwise stated, all the calculations are based
on a system of 9 particles in one dimensional lattice of 100 sites, with an
impurity located at site zero. Periodic boundary condition is imposed on this
one dimensional chain.

\begin{figure}
[ptb]
\begin{center}
\includegraphics[
height=5.3467in,
width=6.5069in
]%
{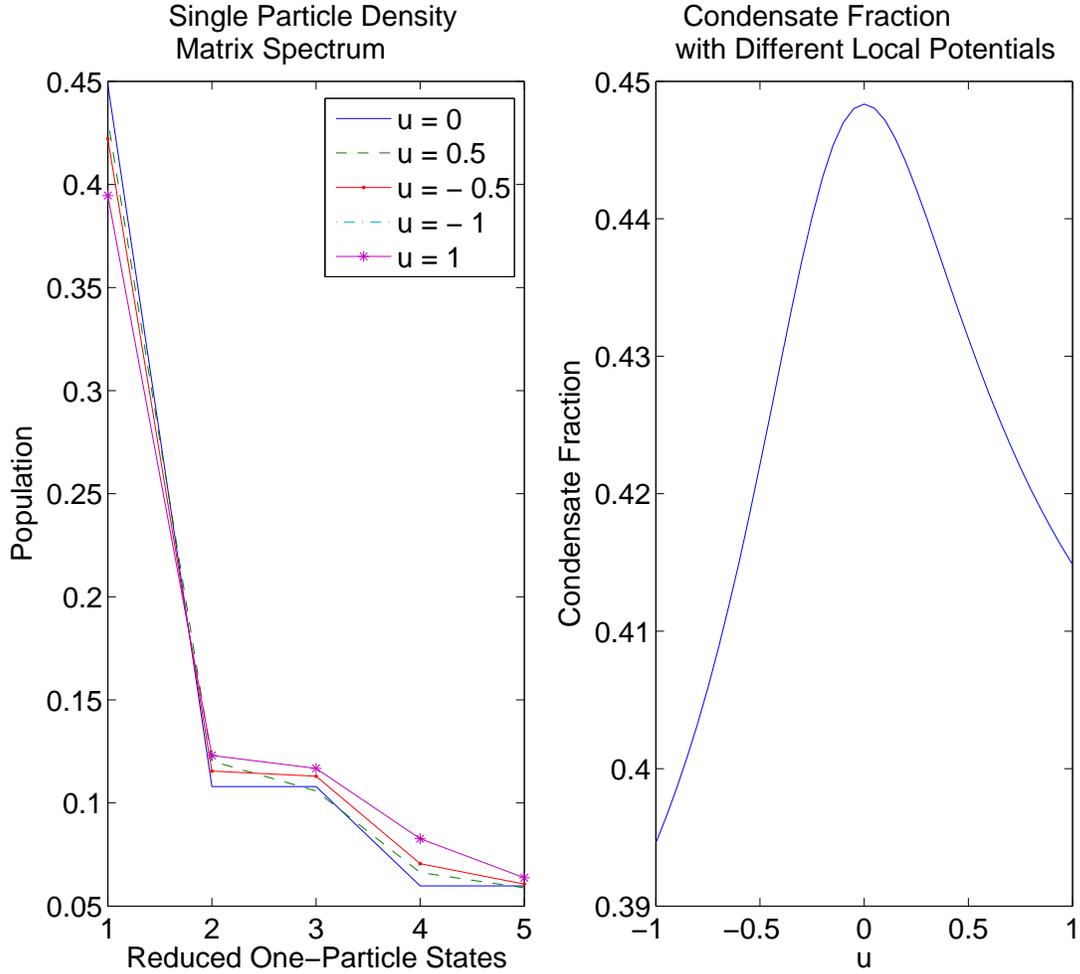}%
\caption{(a) shows the SPDM spectrum at different local potential strength u.
Here we have normalized the population to be one. One of the reduced
one-particle states is \ overwhelming occupied compareing with all the others,
and it is identified as the BEC state. We show only part of spectrum and
omitting parts which are negligibly small. (b) shows the condensate fraction
as a function of the local potential strength. The attractive potential is
seen to have a slightly larger effect in decrease the condensate fraction.}%
\label{SPS}%
\end{center}
\end{figure}

Since all the physics is determined by the ratio $u/t$, in the following
numerical calculations, we fix $t=-1$ and vary $u.$ Fig [\ref{SPS}a] shows
single particle density matrix spectrum for different u's$.$The significant
peak corresponds to the BEC state occupation or condensate fraction. The
introduction of an impurity lowers the condensate fraction, as shown in Fig
[\ref{SPS}b]. The attractive and repulsive potentials almost equally deplete
the condensate fraction for small impurity potential $u$, with the attractive
potential having a slightly larger effect. The physics that gives rise to this
behavior, as is shown in the following, is that an attractive local potential
has a stronger effect on decreasing the coherence between particles.%

\ \ 

\begin{figure}
[ptb]
\begin{center}
\includegraphics[
height=5.3467in,
width=6.5069in
]%
{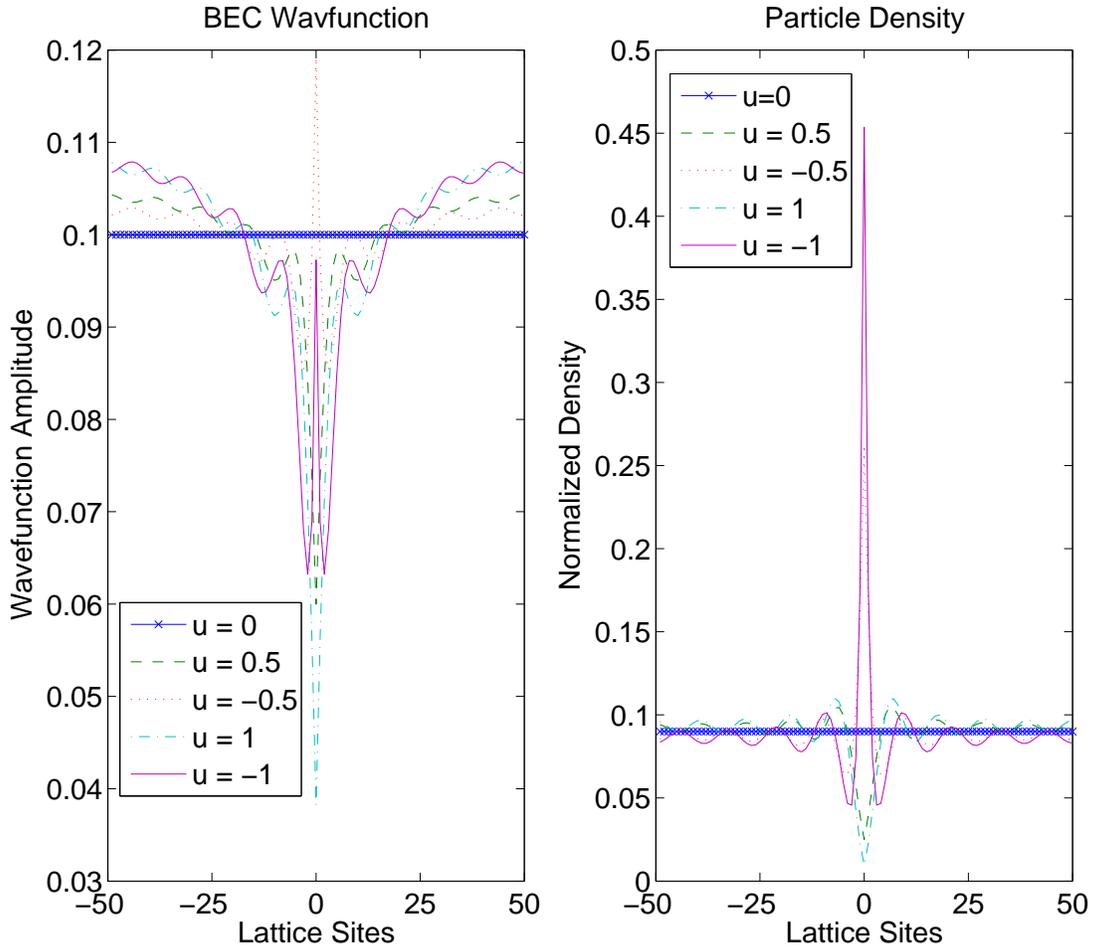}%
\caption{(a) The BEC wavefunctions for different strengthes of the local
potential are shown. In the special case of u=0, the BEC wavefucntion is
constant all over the lattice. Note that the BEC wavfunction corresponding to
$u=-1$ has a lower value at the impurity site comparing with the $u=0$ uniform
case. (b) The particle density is shown with respect to lattice sites for
different u's$.$ Here we see for attractive $u$, the particle density at the
impurity site is always larger than that of the uniform case. }%
\label{MOSPS}%
\end{center}
\end{figure}

\begin{figure}
[ptb]
\begin{center}
\includegraphics[
height=5.3467in,
width=6.5069in
]%
{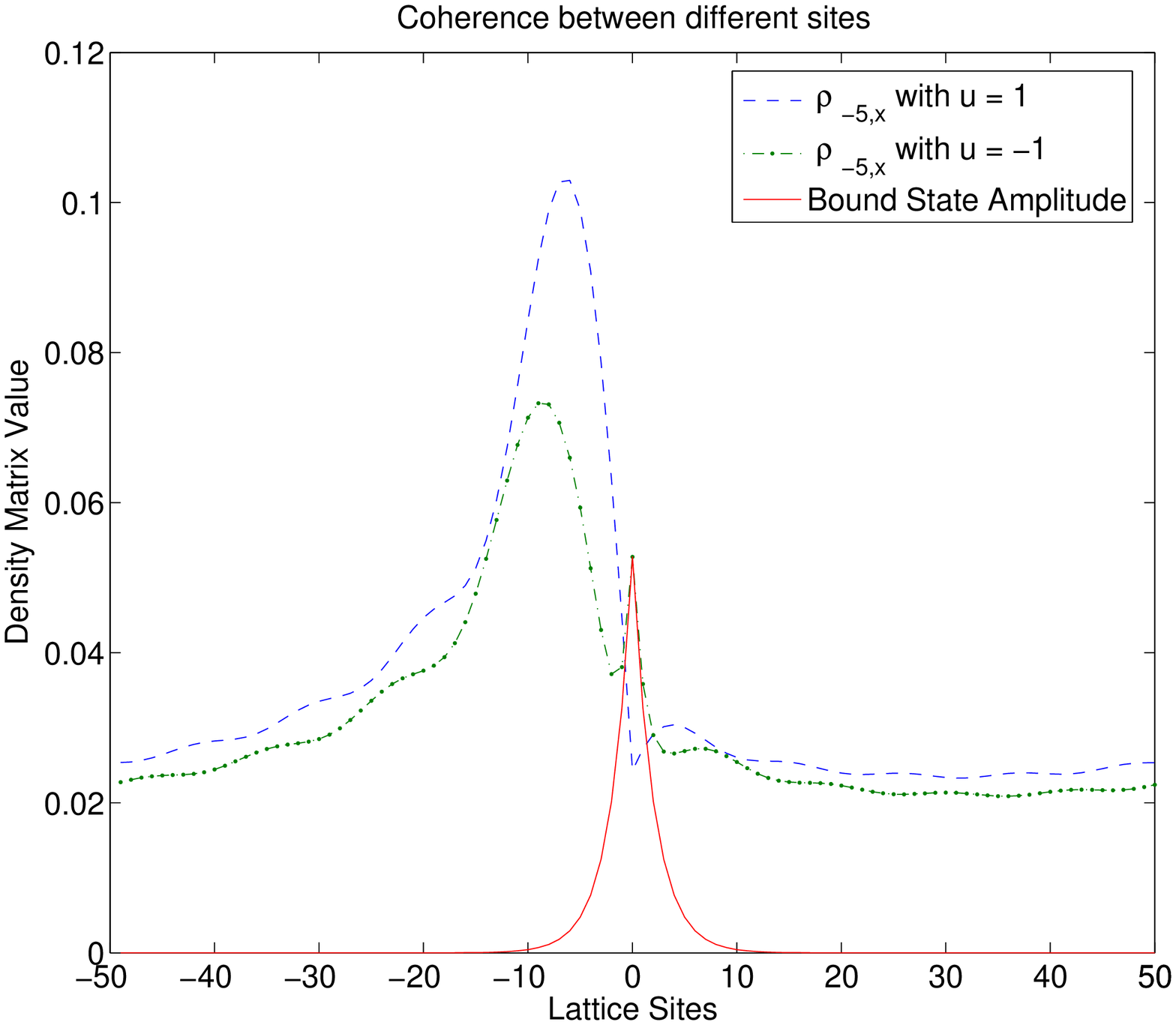}%
\caption{The off diagnal SPDM elements that measure the coherence between
lattice sites -5 and any arbitary lattice sites is plotted for u=-1 and u=1,
respectively. The coherence with the negative sites are relatively larger than
the coherence with the positive sites. The smaller peak of the coherence for
the attractive potential between any sites and the sites zero can be explained
by the increase of the density at impurity site, due to the presence of a
bound state. The bound state wavefunction is plotted and it conincides well
with the peak of the coherence in the impurity site. }%
\label{CO}%
\end{center}
\end{figure}

Since only the BEC state is significantly occupied, the BEC state determines
the most important features of the many-body system. It would be useful to
look at the BEC wavefunction itself(see figure [\ref{MOSPS}a]). For a
repulsive local potential, we find that the BEC density decreases near the
impurity. In the case of an attractive potential, for $u>-1$, there is an
increase in the probability of the BEC density at the impurity site. For the
case of $u<-1$, in contrast to what one might expect, there is a decrease of
the BEC density at the impurity site. This feature actually arises from the
competition between two effects: the single particle effect, i.e. the
potential attracts particles, and the many body effect, i.e. the impurity
decreases the coherence among particles. We have also included the particle
density plot in Fig [\ref{MOSPS}b]. It shows an increase of particle density
for any attractive potential. For the BEC density, we see that, for each
attractive potential, there is a peak at the impurity site corresponding to
the local bound state. However, due to the lack of coherence between bound and
extended states, the bound state is less likely to participate the BEC state,
which results in a overall decrease of the BEC density at site zero compared
with the uniform case. This is the main observation of this paper. To be more
specific on how the impurity decreases the coherence among particles, we plot
some of the relevant off-diagonal elements of the single particle density
matrix (see Fig [\ref{CO}]) In particular, we take the coherence between an
arbitrary site and the fifth site next to the local potential site zero as an
example. We find that, given the same distance, the coherence between
particles on different sides of the impurity is much smaller than that of the
particles located on the same side of the impurity. For the same magnitude of
the attractive and repulsive potential, we find that the effect of the
attractive and repulsive potential are roughly the same, with the attractive
one having a stronger effect in decreasing the overall coherence among
particles. This actually explains the fact the attractive potential has a
stronger effect on decreasing the condensate fraction. The only place that the
attractive potential resulting in a large coherence than the repulsive one
with the same magnitude is near the impurity site. This is due to the presence
of a local bound state, which effectively increase the probability of finding
particles at the impurity site. This bound state is well known \cite{Book} in
the tight binding Hamiltonian. Its eigenenergy is $E=-\sqrt{u^{2}+4t^{2}}$ and
its wavefunction localizes in space as $\exp\left[  -\alpha\left(  E\right)
\left\vert n\right\vert \right]  $ with $n$ being the site number and
\[
\alpha\left(  E\right)  =-\ln\left[  \frac{E}{2t}-\sqrt{\frac{E^{2}}{4t^{2}%
}-1}\right]
\]
corresponds to the inverse of the characteristic length. The nice overlap of
the coherence peak and the bound state wavefunction verifies our argument for
the increase of coherence in the vicinity of impurity. The small peak in the
impurity site is purely a single particle effect.%

%

\begin{figure}
[ptb]
\begin{center}
\includegraphics[
height=5.3467in,
width=6.5069in
]%
{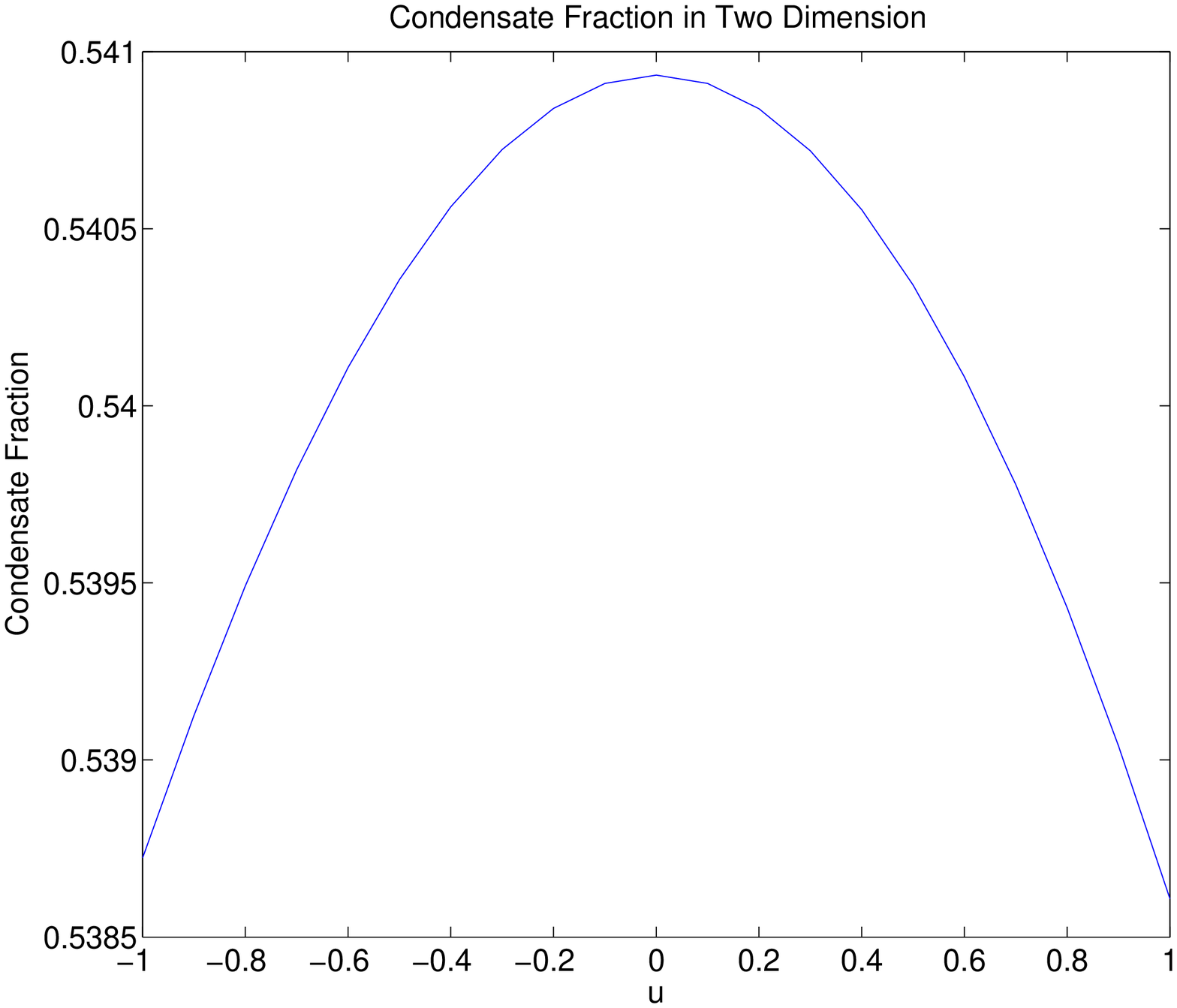}%
\caption{This figure shows the condensate fraction in a three by three lattice
with an impurity located in the center. The impurity is charaterized by a
local potential strength u. The total single particle density spectrum is
normalized to one. We see that in two dimension, the condensate fraction is
larger than its one dimensional counterpart. Morever, the relative change of
the condensate fraction due to the presence of the local potential is much
smaller than the one dimensional case. }%
\label{CF2D}%
\end{center}
\end{figure}

With the picture of the impurity introducing decoherence, one would suggest
that in higher dimensions, since the impurity has a weaker effect in
decreasing coherence among particles, a smaller effect of the impurity should
be found. We have verified this point numerically by extending our calculation
to two dimensions The condensate fraction is shown in Fig[\ref{CF2D}]. At
dimensions greater than one, there is no simple mapping from bosons to
fermions. An exact diagonalization has been done for this case which limits
the calculation to a system of five particles in a three by three lattice.%

\section{Superfluidity}

It is well known that BEC is neither necessary nor sufficient for the
existence of a superfluid. As mentioned by Lieb and Sieringer\cite{Lieb}, a
particular example is the Tonks-Girardeau gas. Even it does not Bose condense,
the system without impurity actually exhibits superfluidity. It is interesting
to investigate how the superfluidity is decreased in the presence of an
impurity. We emphasize that we use the word superfluidity strictly in the
sense of the following phenomenological definition%
\[
\frac{E\left(  v\right)  }{N}-\frac{E\left(  0\right)  }{N}=\frac{1}{2}%
f_{s}mv^{2}+O\left(  v^{4}\right)
\]
where $E\left(  v\right)  $ is the ground state energy of the many body system
in the presence of a perturbative velocity field $v$, $f_{s}$ is the
superfluid fraction, $\ $and $m$ is the mass of the particle. The velocity
field is introduced by imposing a twisted boundary condition \cite{Lieb},
which amounts to a phase jump $e^{i\varphi}$, $\varphi=\frac{vLm}{h}$ whenever
the wave function passes through the boundary. We restrict $\varphi<\pi$ to
yield a single valued function. Note that since the definition is based only
on the static properties of the many body system, it can tell us only whether
the ground state has the property of superfluidity, and it cannot predict the
stability of the superfluidity.

\begin{figure}
[ptb]
\begin{center}
\includegraphics[
height=5.3467in,
width=6.5069in
]%
{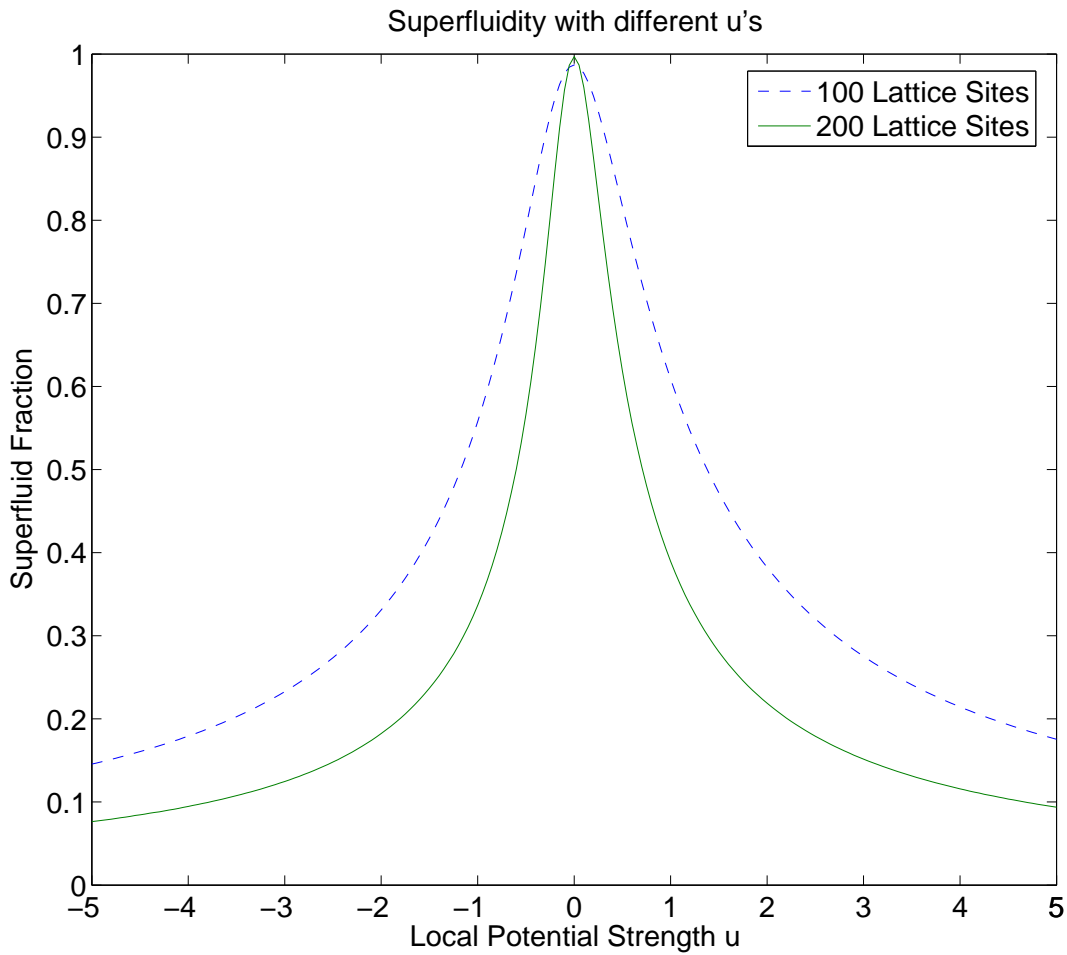}%
\caption{Superfluid fraction is plotted as a function of different local
potential strength u. The case of nine particle occupying 100 and 200 lattice
sites are shown. The larger the system, the sharper the superfuid peak is
found}%
\label{SF}%
\end{center}
\end{figure}

The result of numerical calculations is shown in Fig [\ref{SF}]. We see that
without impurity, the system exhibits superfluidity with superfluid fraction
close to unity. The degree of degradation on the superfluid fraction produced
by impurity depends on both the local impurity strength and the size of the
system. \ In a large system, the superfluidity drops sharply with the presence
of the local impurity. \ Note that as far as the spectrum is concerned, the
fermi ground state is equivalent to the boson ground state. This allows an
understanding of the superfluidity in a single particle picture. As the
lattice size increase, the particle kinetic energy goes like $\frac{1}{L^{2}}$
while the local potential energy goes like $\frac{1}{L}.$ Therefore, in the
limiting case of large $L$, the local potential always dominates the kinetic
energy. For a repulsive potential, the particle has to hop through the
barrier. For an attractive potential, since the local bound state is always
occupied, for hard core bosons, in order to get through this potential, a
potential barrier of order $\frac{1}{L}$ is also present, which gives rise to
the sharp drop of superfluidity in the presence of a local potential.

This problem can be understood quantitatively by considering the following
single particle model $H=-\frac{\hbar^{2}}{2m}\frac{d^{2}}{dx^{2}}%
+g\delta\left(  x\right)  $, with $x$ defined in the regime $\left[
-L,L\right]  $ with the twisted boundary condition. As an example, we consider
only two states with energy close to $\frac{\hbar^{2}\pi^{2}}{2mL^{2}}$. To
calculate the eigenenergy with different boundary condition is elementary. It
is just the one dimensional piecewise potential problem. The modification of
eigenenergy due to the presence of the velocity field is $\left[  E\left(
v\right)  -E\left(  0\right)  \right]  /2=\frac{1}{2}mv^{2}\left(
1-0.008g^{2}L^{2}\right)  $ \cite{SF} in the perturbative limit that $gL$ is
small. This result shows that the superfluid fraction is $f_{s}=1-0.008g^{2}%
L^{2}$. There is no modification to first order in $gL$, which is consistent
with discussions concerning repulsive and attractive potentials.

\section{Discussion and Conclusion}

The effect of an impurity on the Tonks-Girardeau Gas can be studied in an
experiment using an optical lattice. The local potential we model here may be
realized in an experiment by introducing a laser beam, an impurity atom, or an
ion at a particular lattice site. The real space density distribution will be
the same as the free fermion in the presence of a local potential, which
contains only information on the diagonal part of the density matrix. The
non-trivial part of the density matrix, the off diagonal coherence, \ can be
detected by measuring the momentum distribution, defined as
\[
n\left(  k\right)  =\frac{1}{N}\sum_{m,n=-N/2}^{N/2}\rho_{mn}e^{i\left(
m-n\right)  k}%
\]
We have ignored the localized Wannier function profile.

\begin{figure}
[ptb]
\begin{center}
\includegraphics[
height=4.944in,
width=5.6909in
]%
{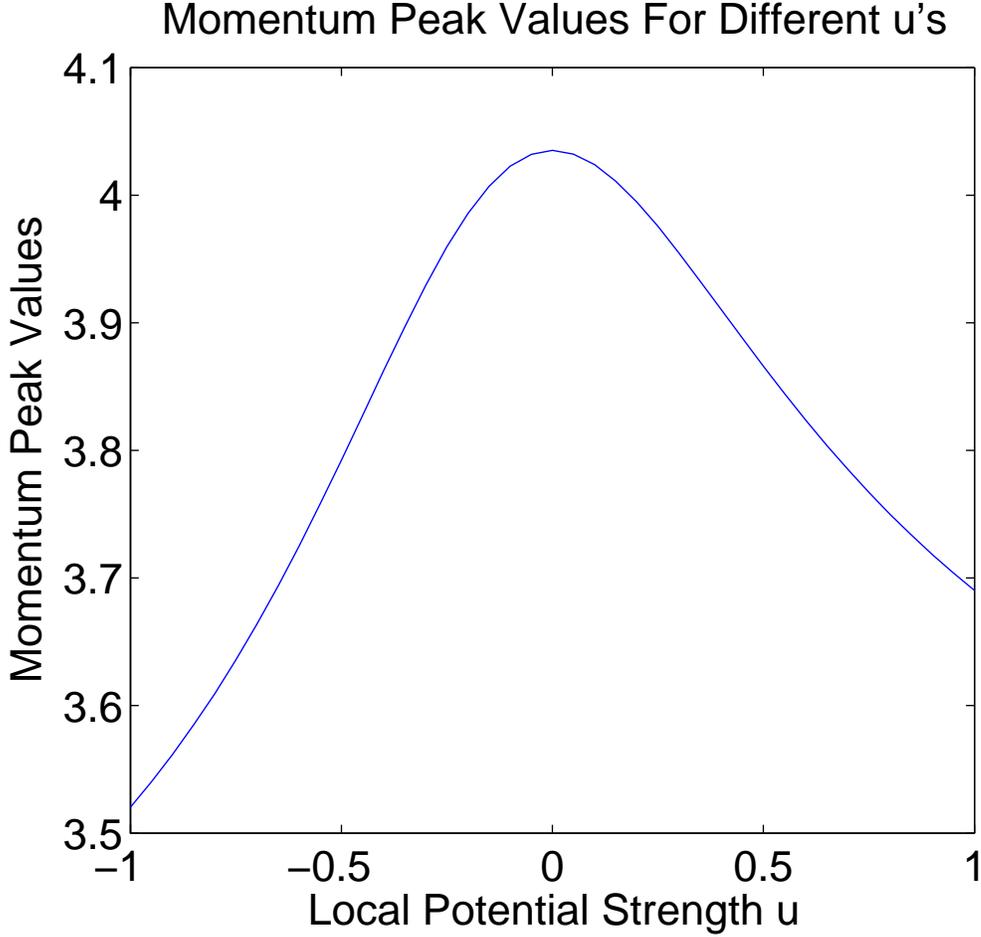}%
\caption{The peak value of the density distribution in the momentum space is
shown. The maximum of zero momentum occupation is reached in the uniform case.
}%
\label{MPV}%
\end{center}
\end{figure}

In figure [\ref{MPV}] we show peak value density in momentum space for
different impurity strengths. The highest momentum density peak corresponds to
the case with no impurity. The attractive potential has a larger effect on the
broadening compared with the repulsive one. The superfluidity of the system
can be measured by imposing a velocity field on the many-body system. This can
be realized in the experiment by changing of the external potential with time.
The measurement can be done by looking at the damping motion of the
particles.

In summary, we have presented a numerical calculation of a Tonks-Girardeau Gas
in the presence of a local potential. It is shown that the impurity decreases
the occupation of condensate fraction. The BEC density at the local potential
decreases in a strong attractive potential, as oppose to particle density,
which always increases for attractive potentials. \ The superfluidity is also
degraded by the impurity, the degradation scales like $g^{2}L^{2}$ in the
small $gL$ limit. This effect of impurity on a many-body system can be
measured by looking at the momentum distribution and the system response an
external velocity field.

\begin{acknowledgments}
We would like to thank P. R. Berman for useful discussions and careful
readings of the manuscript. HF would like to thank supports from NSF and FOCUS
Grant. AR would like to thank supports from NSF and from the Research
Corporation, Cotrell College Science Award.
\end{acknowledgments}


\bigskip

\begin{thebibliography}{99}                                                                                               %


\bibitem {Anderson}P. W. Anderson, Phys. Rev. 109, 1492 (1958); T. Ando and
Fukuyama, \textit{Anderson Localization, }Springer Proceedings in Physics 28
(Springer, Berlin, 1988); A. Auerbach, \textit{Interacting Eletrons and
Quantum Magnetism} (Spinger, New York, 1994) and reference therein.

\bibitem {becexp}Anderson, M. H., J. R. Ensher, M. R. Matthews, C. E. Wieman,
and E. A. Cornell, 1995, Science \textbf{269},198; Davis, K. B., M. -o. Mewes,
M. R. Andrews, N. J. van Druten, D. S. Durfee, D. M. Kurn, and W. Ketterle,
1995, Phys. Rev. Lett. 75, 3969;

\bibitem {Tonksexp}Paredes, B. , Widera, A. , Murg, V., Mandl, O., Folling,
S., Cirac, I. , Shlyapnikov, G. V., Hansch W. T., and Bloch, I., 2004, Nature,
429, 227; T. Kinoshita, T. R. Wenger and D. S. Weiss, 2004, Sicence, 305, 1125

\bibitem {Lenard}A. Lenard, , 1964, J of Math Phys, 5, 930

\bibitem {Lieb}Elliott H. Lieb and Robert Seiringer, 2002, Phys. Rev. B. 66, 134529

\bibitem {Rigol}M. Rigol and A. Muramatsu, 2005, Phys, Rev. A. 72, 013604

\bibitem {IT}R. Cote, V. Kharchenko, and M. D. Lukin, 2002, Phys. Rev. Lett.
89, 093001; P. Horak, J.-Y. Courtois, and G. Grynberg, 1998, Phys. Rev. A 58,
3953; Roberto B. Diener, Georgios A. Georgakis, Jianxin Zhong, Mark Raizen,
and Qian Niu , 2001, Phys. Rev. A 64, 033416; U. Gavish and Y. Castin, 2005,
Phys. Rev. Lett. 95, 020401

\bibitem {IE}J. E. Lye, L. Fallani, M. Modugno, D. S. Wiersma, C. Fort, and M.
Inguscio, 2005, Phys. Rev. Lett. 95, 070401; David Clement, Andres F. Varon,
Mathilde Hugbart, Jocelyn Retter, Philippe Bouyer, Laurent Sanchez-Palencia,
Dimitri M. Gangardt, Georgy V. Shlyapnikov, Alain Aspect, cond-mat/0506638; T.
Schulte, S. Drenkelforth, J. Kruse, W. Ertmer, J. Arlt, K. Sacha, J.
Zakrzewski, M. Lewenstein, cond-mat/0507453

\bibitem {Penrose}Oliver Penrose and Lars Onsager, 1956 Phys. Rev. 104, 576

\bibitem {Book}E. N. Economou, \textit{Green's Functions in Quantum Physics,
1983, }Page 99

\bibitem {SF}Note that in order to solve for the eigenenergy we come up with
the following equation $k\cos\varphi-k\cos(2kL)-\frac{gm\sin(2kL)}{\hbar^{2}%
}=0$ with $k=\sqrt{2mE}/\hbar$. We are looking for solution $k\sim\frac{\pi
}{L}$, this allows a \ expand in order of $k-\frac{\pi}{L}$. This parameter
converges slowly and it is found that one have to expand to $\left(
k-\frac{\pi}{L}\right)  ^{4}$ to yield reasonable result.
\end{thebibliography}
\end{document}